 \documentclass[sigconf]{acmart}
\AtBeginDocument{%
  \providecommand\BibTeX{{%
    \normalfont B\kern-0.5em{\scshape i\kern-0.25em b}\kern-0.8em\TeX}}}

\acmSubmissionID{5866}

\usepackage{enumitem}
\usepackage{multirow, multicol}
\usepackage{graphicx, caption}
\usepackage{array}
\usepackage{makecell}
\usepackage{pifont}
\newcommand{\cmark}{\ding{51}}%
\newcommand{\xmark}{\ding{55}}%
\usepackage{algorithm2e}
\RestyleAlgo{ruled}

\usepackage{amsmath,amsfonts,bbm}

\def\eqref#1{equation~\ref{#1}}

\def\1{\bm{1}}

\def\rvp{{\mathbf{p}}}

\def\rvv{{\mathbf{v}}}

\def\rvx{{\mathbf{x}}}

\def\rmE{{\mathbf{E}}}

\def\rmI{{\mathbf{I}}}

\def\rmP{{\mathbf{P}}}

\def\rmV{{\mathbf{V}}}

\DeclareMathAlphabet{\mathsfit}{\encodingdefault}{\sfdefault}{m}{sl}
\SetMathAlphabet{\mathsfit}{bold}{\encodingdefault}{\sfdefault}{bx}{n}

\def\sV{{\mathbb{V}}}

\newcommand{\R}{\mathbb{R}}

\DeclareMathOperator*{\argmin}{arg\,min}

\usepackage{wasysym}

\begin{document}

\title{Field-wise Embedding Size Search via Structural Hard Auxiliary Mask Pruning for Click-Through Rate Prediction}
\author{Tesi Xiao}
\affiliation{%
  \institution{University of California, Davis}
  \city{Davis}
  \state{CA}
  \country{USA}
}
\email{texiao@ucdavis.edu}

\author{Xia Xiao, Ming Chen, Youlong Chen}
\affiliation{%
  \institution{ByteDance Inc.}
  \city{Mountain View}
  \state{CA}
  \country{USA}
}
\email{{x.xiaxiao, ming.chen, youlong.cheng}@bytedance.com}




\renewcommand{\shortauthors}{Xiao et al.}

\begin{abstract}
Feature embeddings are one of the most essential steps when training deep learning based Click-Through Rate prediction models, which map high-dimensional sparse features to dense embedding vectors. Classic human-crafted embedding size selection methods are shown to be ``sub-optimal" in terms of the trade-off between memory usage and model capacity. The trending methods in Neural Architecture Search (NAS) have demonstrated their efficiency to search for embedding sizes. However, most existing NAS-based works suffer from expensive computational costs, the curse of dimensionality of the search space, and the discrepancy between continuous search space and discrete candidate space. Other works that prune embeddings in an unstructured manner fail to reduce the computational costs explicitly. In this paper, to address those limitations, we propose a novel strategy that searches for the optimal mixed-dimension embedding scheme by structurally pruning a super-net via Hard Auxiliary Mask. Our method aims to directly search candidate models in the discrete space using a simple and efficient gradient-based method. Furthermore, we introduce orthogonal regularity on embedding tables to reduce correlations within embedding columns and enhance representation capacity. Extensive experiments demonstrate it can effectively remove redundant embedding dimensions without great performance loss. 
\end{abstract}





\maketitle

\section{Introduction}

\begin{figure*}[h]
\centering
\includegraphics[width=0.9 \linewidth]{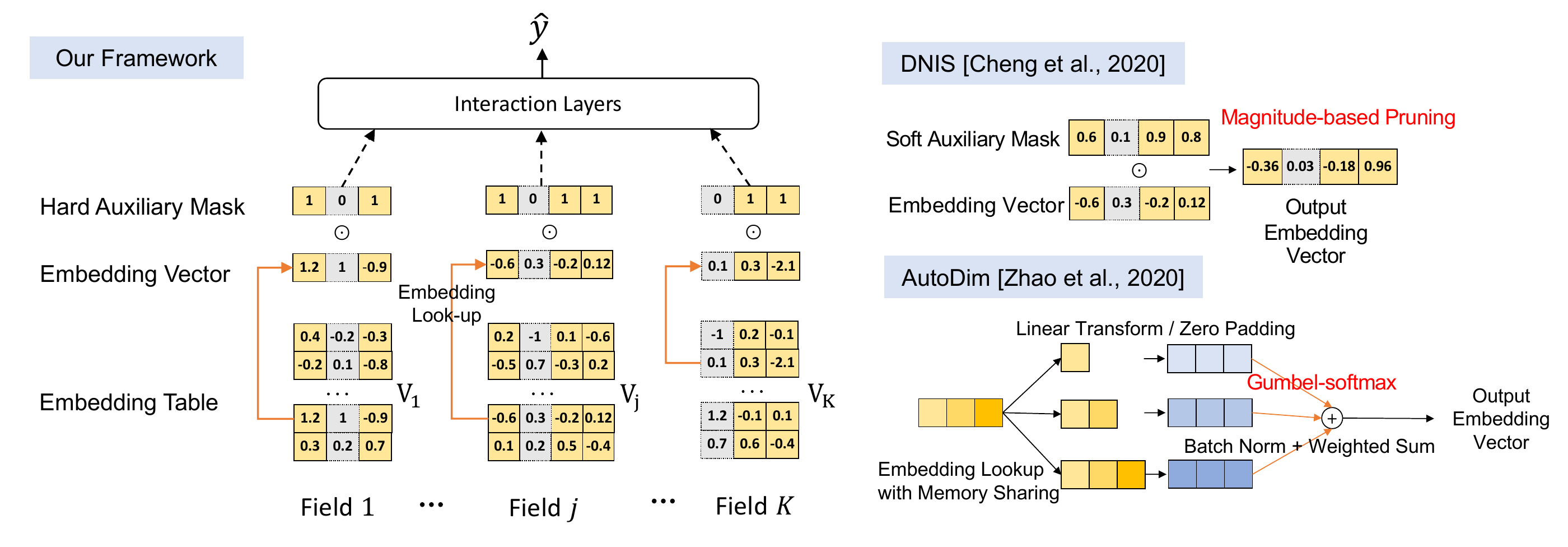}
\caption{ (\textit{left}) The framework of our method. The operator $\odot$ denotes the element-wise product. The gray embedding components are masked and thus can be removed directly. (\textit{right}) The basic ideas of DNIS \citep{cheng2020differentiable} and AutoDim \cite{zhao2020memory}.}
\label{fig:framework}
\end{figure*}

Deep learning based recommender systems (DLRS) have demonstrated superior performance over more traditional recommendation techniques \cite{zhang2019deep}. The success of DLRS is mainly attributed to their ability to learn meaningful representations with categorical features, that subsequently help with modeling the non-linear user-item relationships efficiently. Indeed, real-world recommendation tasks usually involve a large amount of categorical feature fields with high \textit{cardinality}  (i.e. the number of unique values or vocabulary size) \cite{covington2016deep}. \textit{One-Hot Encoding} is a standard way to represent such categorical features. To reduce the memory cost of \textit{One-Hot Encoding}, DLRS first maps the high-dimensional one-hot vectors into real-valued dense vectors via the \textit{embedding} layer. Such embedded vectors are subsequently used in predictive models for obtaining the required recommendations. 

In this pipeline, the choice of the dimension of the embedding vectors, also known as \textit{embedding dimension}, plays a crucial role in the overall performance of the DLRS. Most existing models assign fixed and uniform embedding dimension for all features, either due to the prerequisites of the model input or simply for the sake of convenience. If the embedding dimensions are uniformly high, it leads to increased memory usage and computational cost, as it fails to handle the heterogeneity among different features. As a concrete example, encoding features with few unique values like \textit{gender} with large embedding vectors definitely leads to over-parametrization. 
Conversely, the selected embedding size may be insufficient for highly-predictive features with large cardinality, such as the user’s \textit{last search
query}. Therefore, finding appropriate embedding dimensions for different feature fields is essential. 

The existing works towards automating embedding size search can be categorized into two groups: (i) \textit{field-wise} search \cite{zhao2020autoemb, zhao2020memory, liu2020automated}; (ii) \textit{vocabulary-wise} search \cite{joglekar2020neural,chen2020differentiable,kang2020learning,ginart2021mixed}. The former group aims to assign different embedding sizes to different feature field, while embeddings in the same feature field owns the same dimension. The latter further attempts to find different embedding sizes to different feature values within the same feature field, which is generally based on the frequencies of feature values. Although it has been shown that the latter group can significantly reduce the model size without great performance drop, the second group of works suffer from several challenges and drawbacks (we refer readers to Section 3.4 in \citep{zhao2020memory} for details): (a) the large number of unique values in each feature field leads to a huge search space in which the optimal solution is difficult to find; (b)  the feature values frequencies are time-varying and not pre-known in real-time recommender system; (c) it is difficult to handle embeddings with different dimensions for the same feature field in the state of art DLRS due to the feature crossing mechanism. As a result, we fix our attention to the \textit{field-wise} embedding size search in this work.

The majority of recent works towards automating embedding size search are based on the trending ideas in Neural Architecture Search (NAS). NAS has been an active research area recently, which is composed of three elements: (i) a search space $\mathcal{A}$ of all candidate models; (ii) a search strategy that goes through models in $\mathcal{A}$; (iii) a performance estimation strategy that evaluates the performance of the selected model. To leverage the NAS techniques to search embedding sizes for each feature field, \citet{joglekar2020neural} formulate the search space by discretizing an embedding matrix into several sub-matrices and take a Reinforcement Learning (RL) approach with the \textit{Trainer-Controller} framework for search and performance estimation, in which the controller samples different embedding blocks to maximize the reward observed from the trainer that trains the model with the controller's choices. \citet{liu2020automated} also adopt the RL-based search which starts with small embedding sizes and takes actions to keep or enlarge the size based on the prediction error. Their proposed method requires to sequentially train DRLS and the policy network repeatedly until convergence, which is computationally expensive.

Given the rise of \textit{one-shot NAS} aiming to search and evaluate candidate models together in an over-parametrized supernet, \citet{zhao2020autoemb, zhao2020memory} follow the idea of Differential Neural Architecture Search (DARTS)\citep{liu2018darts} and solve a continuous bilevel optimization by the gradient-based algorithm. To be specific, they define a search space includes all possible embedding size of each feature field; for instance in \cite{zhao2020memory}, 5 candidate sizes $\{2, 8, 16, 24, 32\}$ are selected for each feature field. These embedding vectors in different sizes are then lifted into the same dimension with batch normalization so that they can be aggregated with architecture weights of candidate sizes and fed into the subsequent layers to obtain the prediction. \citet{cheng2020differentiable} also leverage the idea of DARTS but with the architecture weights being the weights of sub-blocks of embedding matrices; see Figure \ref{fig:framework} for further clarifications.

Despite the obtained promising results, the existing methods have certain limitations which arises from the discrepancy between the large discrete candidate search space and its continuous relaxation. On the one hand, the DARTS-based methods in \citep{zhao2020autoemb, zhao2020memory, cheng2020differentiable} is algorithmically efficient but search in an augmented continuous space; it has been shown that the discrepancy may lead to finding unsatisfactory candidate because the magnitude of architecture weights does not necessarily indicate how much the operation contributes to the overall performance \citep{wang2021rethinking}. One the other hand, while the RL-based method in \citep{joglekar2020neural, liu2020automated} and the hard selection version in \citep{zhao2020autoemb} directly search and evaluate models in the discrete candidate space, they are computationally heavy and thus not favored by the large-scale DLRS. Therefore, a natural question follows: \textit{Is it possible to come up with an efficient method that search straight through the discrete candidate space of all possible embedding sizes?}

In this work, we provide a \textbf{positive} answer by proposing an \textit{end-to-end field-aware} embedding size search method leveraging the Straight Through Estimator (STE) of gradients for non-differentiable functions. Motivated by one-shot NAS and network pruning, our method seek to adaptively search for a good sub-network in a pretrained supernet by masking redundant dimensions. The mask, named as \textbf{Hard Auxiliary Mask} (HAM), is realized by the indicator function of auxiliary weights. Furthermore, to reduce the information redundancy and boost the search performance, we impose the orthogonal regularity on embedding tables and train models with regularized loss functions. \textbf{Contributions and novelty} of our work are: (i) we propose to prune embedding tables column-wisely via hard auxiliary masks for the CTR prediction models, which can effectively compress the model; (ii) we offer a gradient-based pruning method and delve into its dynamics with mathematical insights; (iii) we introduce orthogonal regularity for training recommendation models and experimentally show that orthogonal regularity can help to achieve significant improvement; (iv) we obtain state-of-art results on various modern models and our method is scalable on large datasets.

\section{Related Work}

\textbf{Embedding Size Search for Recommender Systems.} To deal with the heterogeneity among different features and reduce the memory cost of sizable embedding tables, several recent works introduce a new paradigm of mixed-dimension embedding table. In particular, \citet{ginart2021mixed} propose to substitute the uniform-dimension embeddings $\rmV\in \R^{C\times d}$ with $\overline{\rmV} = \rmE\rmP$, where $\rmE\in\R^{C\times s}$ is the smaller embeddings with $s\ll d$ and $\rmP\in\R^{s\times d}$ is the projection that lift the embeddings into the base dimension $d$ for feature crossing purposes. A popularity-based rule is further designed to find the dimension $s$ for each feature field which is too rough and cannot handle important features with low frequency. In addition, plenty of works seek to provide an end-to-end framework leveraging the advances in NAS, including RL approaches \citep{joglekar2020neural, liu2020automated} and differentiable NAS methods \citep{zhao2020autoemb, bender2018understanding, zhao2020memory, cheng2020differentiable}. The most relevant work to us is \cite{cheng2020differentiable}, which employs a soft auxiliary mask to prune the embeddings in a structural way. However, it requires magnitude-based pruning to deriving fine-grained mixed embedding, in which the discrepancy between the continuous relaxed search space and discrete candidate space could lead to relatively great loss in performance. The most recent work \cite{liu2021learnable} proposes a Plug-in Embedding Pruning (PEP) approach to prune embedding tables into sparse matrices to reduce storage costs, Albeit significantly reducing the number of non-zero embedding parameters, this type of unstructured pruning method fails to explicitly reduce the embedding size.

\textbf{Neural Network Pruning via Auxiliary Masks.} To deploy deep learning models on resource-limited devices, there is no lack of works in neural network pruning; see \cite{blalock2020state} and the reference therein. One popular approach towards solving this problem is to learn a pruning mask through auxiliary parameters, which considers pruning as an optimization problem that tends to minimize the supervised loss of masked network with certain sparsity constraint. As learning the optimal mask is indeed a discrete optimization problem for binary inputs, existing works attempt to cast it into a differentiable problem and provide gradient-based search algorithms. A straightforward method is to directly replace binary values by smooth functions of auxiliary parameters during the forward pass, such as sigmoid \cite{savarese2019winning}, Gumbel-sigmoid \cite{louizos2018learning}, softmax \cite{liu2018darts}, Gumbel-softmax \cite{xie2018snas}, and piece-wise linear \cite{guo2016dynamic}. However, all these methods, which we name as Soft Auxiliary Mask (SAM), suffer from the discrepancy caused by continuous relaxation. Other methods, which we refer to Hard Auxiliary Mask (HAM), preserve binary values in the forward pass with indicator functions \cite{xiao2019autoprune} or Bernoulli random variables \cite{zhou2021effective} and optimize parameters using the straight-through estimator (STE) \cite{bengio2013estimating, jang2016categorical}. A detailed comparison of four representative auxiliary masking methods for approximating binary values is provided in Table \ref{tab:mask}. 

\begin{table*}
\small
  \caption{A List of Representative Auxiliary Masking Methods}
  \label{tab:mask}
  \begin{tabular}{|c|c|c|c|c|}
    \hline
    \multicolumn{2}{|c|}{\textbf{Mask}}  & \textbf{Forward} & \textbf{Backward} & \textbf{$\text{Model}_{\text{eval}}=\text{Model}_{\text{sel}}$} \\
    \hline
    \multirow{2}{*}{Soft} & Deterministic & \makecell{$\alpha\in[0,1]$\cite{cheng2020differentiable}; $\text{Sigmoid}(\alpha), \alpha\in \R$\cite{savarese2019winning}} & autograd &\xmark \\ 
    \cline{2-5}
    & {Stochastic}&  \makecell{$\text{Sigmoid}(\log\alpha + \log(\frac{u}{1-u}))$ \cite{louizos2018learning},\\ $\alpha\in\R, u\sim \text{Uniform(0,1)}$} & autograd & \xmark \\
    \hline
    \multirow{2}{*}{Hard} & Stochastic & Bernoulli(p) & \makecell{STE \cite{srinivas2017training}, Gumbel-STE \cite{zhou2021effective}} & \xmark \\
    \cline{2-5}
    &Deterministic & \makecell{ $\mathbf{1}_{\alpha>0}$ }& STE \cite{xiao2019autoprune, ye2020adaptive} & (Our Approach) \cmark \\
    \hline
  \end{tabular}
  \vskip 1em
  \textbf{Remark.} $\text{Model}_{\text{eval}}$ \textit{ stands for the masked model evaluated by the algorithm;} $\text{Model}_{\text{sel}}$ \textit{ denotes the model finally selected by the algorithm.} 
\end{table*}

\section{Methodology}

\subsection{Preliminaries}

Here we briefly introduce the mechanism of DLRS and define terms used throughout the paper.

\textbf{Model Architecture.} Consider the data input involves $K$ features fields from users, items, their interactions, and contextual information. We denote these raw features by multiple one-hot vectors $\rvx_1\in\R^{C_1}, \dots, \rvx_K\in\R^{C_K}$, where field dimensions $C_1,\dots C_K$ are the cardinalities of feature fields\footnote{Numeric features are converted into categorical data by binning.}.  Architectures of DLRS often consists of three key components: (i) \textit{embedding} layers with tables 
$\rmV_1\in\R^{C_1\times d_1}, \dots, \rmV_K\in\R^{C_K\times d_K}$ that map sparse one-hot vectors to dense vectors in a low dimensional embedding space by $\rvv_i = \rmV_i^\top \rvx_i$; (ii) \textit{feature interaction} layers that model complex feature crossing using embedding vectors; (iii) \textit{output} layers that make final predictions for specific recommendation tasks. 

The feature crossing techniques of existing models fall into two types - \textit{vector-wise} and \textit{bit-wise}. Models with \textit{vector-wise} crossing explicitly introduce interactions by the inner product, such as Factorization Machine (FM) \cite{rendle2010factorization}, DeepFM \cite{guo2017deepfm} and AutoInt \cite{song2019autoint}. The \textit{bit-wise} crossing, in contrast, implicitly adds interaction terms by element-wise operations, such as the outer product in Deep Cross Network (DCN) \cite{wang2017deep}, and the Hadamard product in NFM \cite{he2017neural} and DCN-V2 \cite{wang2021dcn}. In this work, we deploy our framework to the Click-Through Rate (CTR) prediction problem with four base models: FM, DeepFM, AutoInt, DCN-V2.

\textbf{Notations.} Throughout this paper, we use $\odot$ for the Hadamard (element-wise) product of two vectors. The indicator function of a scalar-valued $\alpha $ is defined as $\mathbf{1}_{\alpha>0} = 1 \text{ if } \alpha>0; \mathbf{1}_{\alpha>0}=0 \text{ if } \alpha\leq 0$. The indicator function of a vector $\boldsymbol{\alpha}\in\mathbb{R}^d$ is defined as $\mathbbm{1}_{\boldsymbol{\alpha}>0} = [\mathbf{1}_{\alpha_1>0},\dots, \mathbf{1}_{\alpha_d>0}]^\top$. The identity matrix is written as $\mathbf{I}$ and the function $\text{diag}(\cdot)$ returns a diagonal matrix with its diagonal entries as the input. We use $\|\cdot\|_1, \|\cdot\|_F$ for the $\ell_1$ norm of vectors and the Frobenius norm of matrices respectively.

\subsection{Background}

In general, the CTR prediction models takes the concatenation of all feature fields from a user-item pair, denoted by $\rvx=[\rvx_1; \rvx_2; \dots; \rvx_K]$ as the input vector. Given the embedding layer $\sV = \{\rmV_1, \rmV_2, \dots, \rmV_K\},$ the feature interaction layers take the embedding vectors $\rvv$ and feed the learned hidden states into the output layer to obtain the prediction. The embedding vectors $\rvv$ are the dense encoding of the one-hot input $\rvx$ that can be defined as follows:
\begin{equation*}
\rvv = \left[\rvv_1; \rvv_2; \dots; \rvv_K\right] = \left[\rmV_1^\top \rvx_1; \rmV_2^\top \rvx_2; \dots; \rmV_K^\top \rvx_K\right] := \mathcal{V} \rvx,
\end{equation*}
where $\mathcal{V}$ is the embedding look-up operator. The prediction score $\hat{y}$ is then calculated with models' other parameters $\boldsymbol{\Theta}$ in the feature interaction layers and output layer by $
\hat{y} = \psi(\rvv | \boldsymbol{\Theta}) =  \psi( \mathcal{V} \rvx  | \boldsymbol{\Theta}) = \phi (\rvx | \sV, \boldsymbol{\Theta}),    
$
where $\hat{y}\in [0,1]$ is the predicted probability, $\psi(\cdot | \Theta)$ is the prediction function of embedding vectors, and $\phi = \psi \circ \mathcal{V}$ is the prediction function of raw inputs. To learn the model parameters $\sV, \boldsymbol{\Theta}$, we aim to minimize the Log-loss on the training data, i.e.,
\begin{equation*}
    \min_{\sV, \boldsymbol{\Theta}}~ \mathcal{L}_{\text{train}} (\sV, \boldsymbol{\Theta}) := - \frac{1}{N} \sum_{j=1}^N \left(y_j \log (\hat{y}_j) + (1-y_j) \log(1-\hat{y}_j)\right)
\end{equation*}
where $N$ is the total number of training samples.

\textbf{Hard Auxiliary Mask} As is illustrated in Figure \ref{fig:framework}, we add auxiliary masks for each embedding dimension slots. Specifically, the auxiliary masks are indicator functions of auxiliary parameters $\boldsymbol{\alpha} = [\boldsymbol{\alpha_1}; \boldsymbol{\alpha}_2; \dots; \boldsymbol{\alpha}_K]$, where $\boldsymbol{\alpha}_i\in\R^{d_i}$ is in the same size of the corresponding embedding vector $\rvv_i$. Provided with the masks, the predicted probability score $\Tilde{y}$ is given as follows:
\begin{equation*}
    \Tilde{y} = \psi(\rvv \odot \mathbbm{1}_{\boldsymbol{\alpha}>0}| \boldsymbol{\Theta}) = \phi (\rvx | \Tilde{\sV}_{\boldsymbol{\alpha}}, \boldsymbol{\Theta}) =   \phi (\rvx | \boldsymbol{\alpha}, \sV, \boldsymbol{\Theta}),
\end{equation*}
where $\Tilde{\sV}_{\boldsymbol{\alpha}} = \{\Tilde{\rmV}_1^{\boldsymbol{\alpha}_1},\dots, \Tilde{\rmV}_K^{\boldsymbol{\alpha}_K} \} $ is the pruned embedding layer with $$\Tilde{\rmV}_i^{\boldsymbol{\alpha}_i} = \rmV_i ~\text{diag}(\mathbbm{1}_{\boldsymbol{\alpha}_i>0}), \quad i=1,\dots K.$$ We emphasize that embedding tables $\Tilde{\rmV}_i^{\boldsymbol{\alpha}_i}$ are pruned column-wisely in a structural manner unlike PEP \cite{liu2021learnable} that prunes entry-wisely.

\textbf{Orthogonal Regularity} Given the embedding table $\rmV_j\in\R^{C_j\times d_j}$ for feature field $j$, its column vectors, denoted by $ \rmV_{j,1}, \dots, \rmV_{j,d_j} \in\R^{C_j},$ can be regarded as $d_j$ different representations of feature $j$ in the embedding space. The auxiliary masks above aim to mask relatively uninformative column vectors so as to reduce the model size. Nonetheless, the presence of correlation between these vectors may complicate the selection procedure. Specifically, presuming that the most predictive one $\rmV_{j, p}$ has been selected, it would be problematic if we greedily select the next column $\rmV_{j, q}$ that brings the largest loss drop when included in. For instance, if $\rmV_{j, q} \not\perp \rmV_{j, p}$, we have the following decomposition:   $ \rmV_{j, q} = \rvp + \rvp^{\perp},$ where $\rvp = c \rmV_{j, p} \parallel \rmV_{j, p}$ and $\rvp^{\perp}\perp \rvp$. Therefore, it would be difficult to determine whether the increments are attributed to the existing direction $\rvp$ or the new factor $\rvp^\perp$. To address this issue, we follow \cite{bansal2018can} to train embedding parameters $\sV$ with Soft Orthogonal (SO) regularizations:
\begin{equation}\label{eq:SO}
    \mathcal{R}(\sV) = \sum_{j=1}^{K} ~\|  \rmV_j^\top \rmV_j  - \rmI \|_F^2 / d_j^2,
\end{equation}
where divisors $d_j^2$ are introduced to handle heterogeneous dimensionality of embedding tables. We also adopt a relaxed SO regularization in which $\rmV_j$ replaced by the normalized matrix $\overline{\rmV}_j$ with unit column vectors, which corresponds to the pair-wise cosine similarities within embedding columns \cite{rodriguez2016regularizing}.

\subsection{Framework}

Our proposed framework is motivated by one-shot NAS, which consists of three stages: \textit{pretrain}, \textit{search}, and \textit{retrain}.

\textbf{Pretrain.} As shown in \citep{yan2020does}, pre-training architecture representations improves the downstream architecture
search efficiency. Therefore, in the pretraining stage, we train the base model with a large embedding layer $\sV$. The base dimension $d_j$ for each feature field are determined by prior knowledge. In addition, the embedding dimension $d_j$ should not exceed the field dimension $C_j$ to avoid column-rank-deficiency. The SO regularization term (\ref{eq:SO}) is added to mini-batch training loss for optimizer to learn near-orthogonal embeddings. The learned model parameters, denoted by $\sV_{\text{init}}, \boldsymbol{\Theta}_{\text{init}}$, are passed to the search stage as initialization.

\textbf{Search.} Provided with the pre-trained model, the goal of search stage is to find the column-wise sparse embedding layer that preserves model accuracy, which can be formulated as:
\begin{equation*}
    \underset{\boldsymbol{\alpha}}{\min} ~\underset{\sV, \boldsymbol{\Theta}}{\min} \quad \mathcal{L}_{\text{train}}\left( \Tilde{\sV}_{\boldsymbol{\alpha}}, \boldsymbol{\Theta}  \right) +\mu \big\vert \| \mathbbm{1}_{\boldsymbol{\alpha}>0} \|_1 - s \big\vert,
\end{equation*}
where $\| \mathbbm{1}_{\boldsymbol{\alpha}>0}  \|_1$ counts the number of non-zero embedding columns and  $s$ is the target number of non-zero columns. Note that instead of direct regularization on $\| \mathbbm{1}_{\boldsymbol{\alpha}>0}  \|_1$ as \cite{savarese2019winning, xiao2019autoprune, ye2020adaptive} do, we include the target number $s$ to reduce instability from batched training and the choice of hyperparameter $\mu$. However, given that the objective function above is non-differentiable when $ \alpha = 0$ and has zero gradient anywhere else, traditional gradient descent methods are not applicable. To that end, we use
the straight-through estimator (STE) \cite{bengio2013estimating}, which replaces the ill-defined gradient in the chain rule by a fake gradient. In spite of various smooth alternative used in the literature, such as sigmoid \cite{savarese2019winning} and piecewise polynomials \cite{hazimeh2021dselect}, we adopt the simplest identity function for backpropagation, i.e., 
\begin{equation}\label{eq:STE-gradient}
    \frac{\partial \mathcal{L}}{\partial \boldsymbol{\alpha}} = \frac{\partial \mathcal{L}}{\partial \mathbbm{1}_{\boldsymbol{\alpha}>0}}\frac{\partial \mathbbm{1}_{\boldsymbol{\alpha}>0}}{\partial \boldsymbol{\alpha}}\approx \frac{\partial \mathcal{L}}{\partial \mathbbm{1}_{\boldsymbol{\alpha}>0}}\frac{\partial \boldsymbol{\alpha} }{\partial \boldsymbol{\alpha}} = \frac{\partial \mathcal{L}}{\partial \mathbbm{1}_{\boldsymbol{\alpha}>0}}.
\end{equation}
Then, the search stage starts with the unmasked model that $\boldsymbol{\alpha}_0 = \epsilon \cdot \vec{1}$ for some small $\epsilon>0$. The gradient update rules for $\boldsymbol{\alpha}$ at iteration $t$ are given by:
\begin{equation}\label{eq:gradient-updata-alpha}
    \boldsymbol{\alpha}_{t+1} =  \boldsymbol{\alpha}_{t} -  \eta \cdot \nabla_{(\mathbbm{1}_{\boldsymbol{\alpha}_t >0})} \mathcal{L}_{\text{batch}} - \mu \cdot \text{sign}(\| \mathbbm{1}_{\boldsymbol{\alpha}_t >0} \|_1 - s)  \vec{1},
\end{equation}
where $\eta$ is the learning rate. We will illustrate in Section \ref{sec:gradient-dynamics} that the above updates enjoy certain benefits and the last term plays an important role by pushing the optimizer iteratively evaluating candidate models with hard auxiliary mask. Furthermore, to enhance the stability and performance, we implement a multi-step training through iteratively training auxiliary parameters on validation data and retraining the original weights, which attempts to solve the following bi-level optimization problem:
\begin{equation*}
    \begin{split}
        \underset{\boldsymbol{\alpha}}{\min}\quad &\mathcal{L}_{\text{val}}\left( \Tilde{\sV}_{\boldsymbol{\alpha}}^\star, \boldsymbol{\Theta}^\star  \right) + \mu \big\vert \| \mathbbm{1}_{\boldsymbol{\alpha}>0} \|_1 - s \big\vert\\
        \text{s.t.}\quad & \Tilde{\sV}_{\boldsymbol{\alpha}}^\star, \boldsymbol{\Theta}^\star = \underset{\Tilde{\sV}_{\boldsymbol{\alpha}}, \boldsymbol{\Theta} }{\argmin}\quad \mathcal{L}_{\text{train}}\left( \Tilde{\sV}_{\boldsymbol{\alpha}}, \boldsymbol{\Theta} \right).\\
    \end{split}
\end{equation*}

\textbf{Early Stopper for the Search Stage.}\label{sec:search-early-stopper} We can determine to stop the search stage if the number of positive auxiliary weights is close to the target size with no significant gain in validation AUC, since the sign of auxiliary weights exactly indicates whether the corresponding embedding components are pruned or not. On the contrary, it is hard to deploy the early stopper for the search stage using other auxiliary mask pruning methods because the value of validation AUC during the search epochs is unable to represent the performance of the model selected in the retraining stage.


\textbf{Retrain.} After training $\{\boldsymbol{\alpha}, \sV, \boldsymbol{\Theta}\}$ for several epochs in the search step, we obtain a binary mask based on the sign of $\boldsymbol{\alpha}$ and continue optimizing $\{ \sV, \boldsymbol{\Theta}\}$ till convergence. The early stopper is deployed for both pretraining and retraining steps that terminates training if model performance on validation data could not be improved within a few epochs. The overall framework is described in Algorithm \ref{alg:framework}.
\begin{algorithm}
\caption{Structural HAM Pruning}
\label{alg:framework}
 \textbf{Input:} data $\mathcal{D}_{\text{train}}, \mathcal{D}_{\text{val}}, \mathcal{D}_{\text{test}}$,\\
 base dimensions $d_j(\leq C_j$) for each field $j$,\\ target embedding size $s$, hyperparameters $\mu,\epsilon>0$\;
 $\triangleright$ \textbf{Pretrain}:\\
 \While{stopping criteria not met}{
 obtain a batch of samples from $\mathcal{D}_{\text{train}}$ and update $\sV, \boldsymbol{\Theta}$ regularized with SO (\ref{eq:SO}) by certain optimizer\;}
 $\triangleright$ \textbf{Search}:
 initialize $\boldsymbol{\alpha} = \epsilon \vec{1}$\;
 \While{stopping criteria not met}{
  obtain a batch of samples from $\mathcal{D}_{\text{val}}$ and update $\boldsymbol{\alpha}$ by Eq. (\ref{eq:gradient-updata-alpha})\;
  obtain a batch of samples from $\mathcal{D}_{\text{train}}$ and update $\sV, \boldsymbol{\Theta}$ by certain optimizer\;
 }
 $\triangleright$\textbf{Retrain}: mask the dimensions with 0 where $\boldsymbol{\alpha}<0$ and retrain the model until the stopping criteria are met.
\end{algorithm}

\subsection{On the Gradient Dynamics}\label{sec:gradient-dynamics}

As finding the optimal mask is essentially a combinatorial optimization problem over the set of $2^{S}$ possible status of on-off switches ($S$ is the number of switches), the proposed gradient-based search rule given in (\ref{eq:gradient-updata-alpha}) provides an alternative approach to look through the discrete candidate sets in a continuous space. We elaborate below that the STE-based gradient in (\ref{eq:STE-gradient}) is related to the Taylor approximation for the Leave-One-Out error and the penalty term drifts auxiliary variables to find a model of the target size.

For illustration purpose, we start with the dynamics of the scalar-valued parameter $\alpha_{i,j}$, the $j$-th element of $\boldsymbol{\alpha}_i$, that controls the mask for the $j$-th column vector, $\rmV_{i,j}$, of the embedding table $\rmV_{i}$. Define the function $\ell(c):=\mathcal{L}_{[c\cdot\rmV_{i,j}]}$ as the value of loss function when replacing $\rmV_{i,j}$ by $c\cdot \rmV_{i,j}$ ($0\leq c\leq1$). By the first-order Taylor approximation, we have $ \ell(1) - \ell(0)  \approx \ell'(0)$ and $ \ell(0) - \ell(1) \approx -\ell'(1)$. Moreover, it is worth noting that the STE-based gradient in (\ref{eq:STE-gradient}) with regards to $\alpha_{i,j}$ is exactly $\ell'(\mathbf{1}_{\alpha_{i,j} > 0})$, i.e.,
\begin{equation}\label{eq:Taylor-approx}
    \frac{\partial \mathcal{L}_{\text{batch}}}{\partial \mathbf{1}_{\alpha_{i,j}>0}}  = \ell'(\mathbf{1}_{\alpha_{i,j}>0}) \approx \ell(1) - \ell(0) = \mathcal{L}_{[\rmV_{i,j}]} - \mathcal{L}_{[0\cdot\rmV_{i,j}]}.
\end{equation}
In other words, the proposed gradient calculates the importance (with batches of data) of the embedding component $\rmV_{i,j}$ using the first-order Taylor approximation. The above heuristics are also mentioned in \citep{molchanov2016pruning, molchanov2019importance, ye2020adaptive}.

\begin{figure}[h]
\begin{center}
\includegraphics[width=0.8\linewidth]{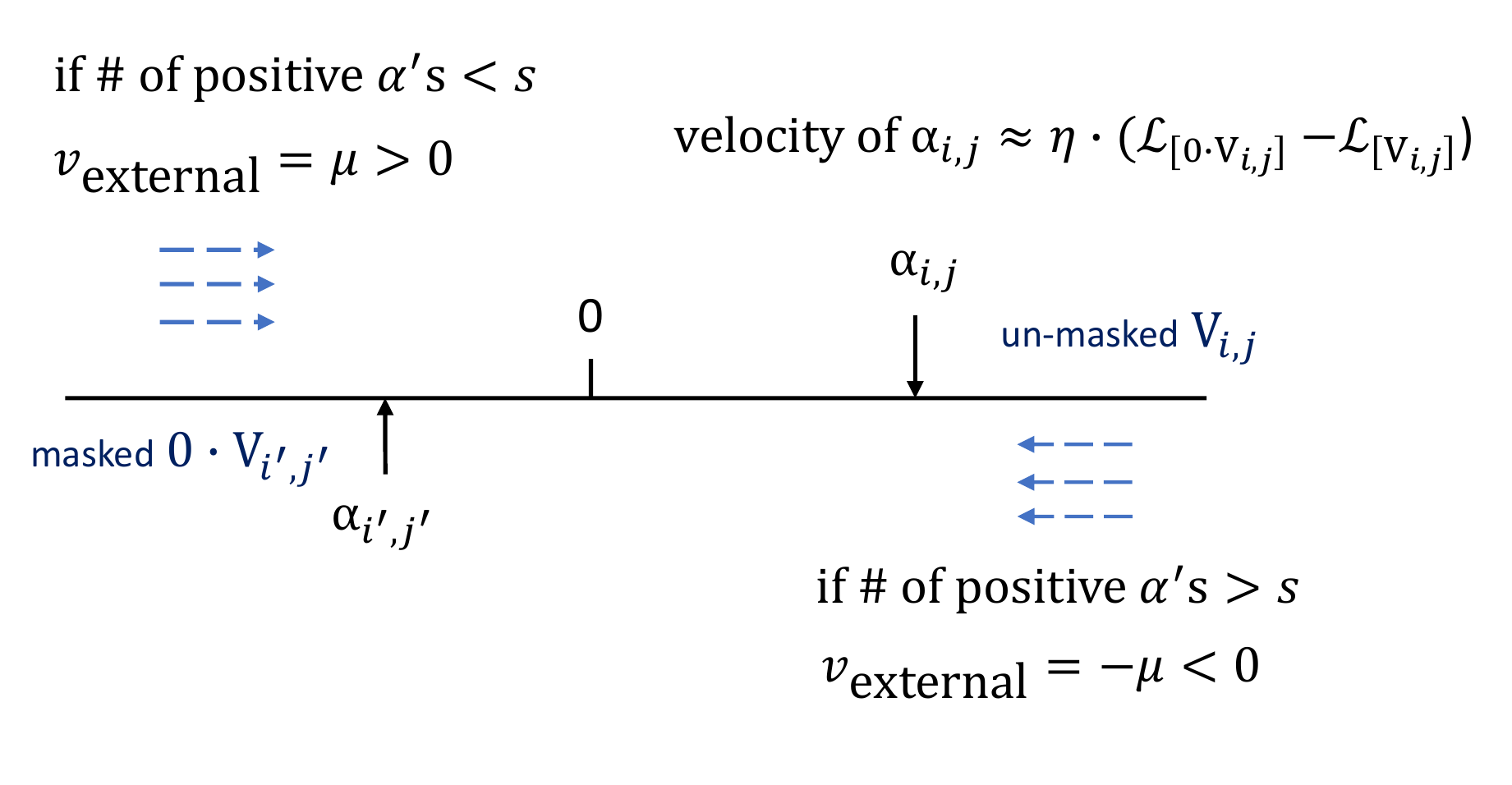}
\end{center}
\vskip -1em
\caption{An intuitive viewpoint of the gradient dynamics.}
\label{fig:dynamics}
\end{figure}

We now consider the dynamics of all the auxiliary parameters $\alpha_{i,j}$'s provided by (\ref{eq:gradient-updata-alpha}). As illustrated above, the term $\nabla_{(\mathbbm{1}_{\boldsymbol{\alpha}_t >0})} \mathcal{L}_{\text{data}} $ measures the importance of each embedding component by approximating the difference between the loss value with un-masked embeddings and the loss value with a mask on. Furthermore, the sign of the penalty term $\mu$ in (\ref{eq:gradient-updata-alpha}) is determined by the comparison between the number of un-masked components and the target size $s$. As an analogue, the dynamics of auxiliary parameters can be viewed as a particle system in a neighborhood of 0 on real line, in which the particles are initialized at the same point $\epsilon>0$ (i.e., the model starts with all embeddings un-masked) and the velocity of each particle is roughly $\eta\cdot (\mathcal{L}_{[0\cdot\rmV_{i,j}]} - \mathcal{L}_{[\rmV_{i,j}]})$ by the approximation in (\ref{eq:Taylor-approx}). In addition, an external force is introduced by the penalty term, $\mu \big\vert \| \mathbbm{1}_{\boldsymbol{\alpha}>0} \|_1 - s \big\vert$, which can be interpreted as the wind flow with its velocity being $-\mu<0$ when the number of positive particles are larger than $s$ and being $\mu>0$ otherwise. As a result, when the number of positive particles exceeds $s$, those particles with smaller $\mathcal{L}_{[0\cdot\rmV_{i,j}]} - \mathcal{L}_{[\rmV_{i,j}]}$ tend to be negative, and vice versa; see Figure \ref{fig:dynamics} for further illustration of the proposed algorithm.


\section{Experiments}

To validate the performance of our proposed framework, we conduct extensive experiments on three real-world recommendation datasets. Through the experiments, we seek to answer the following three questions: (i) How does our proposed method, \texttt{HAM}, perform compared with other auxiliary mask pruning (AMP) methods? (ii) How does our proposed framework perform compared with the existing field-wise embedding size search methods in the literature in terms of prediction ability, memory consumption? (iii) How does the soft orthogonality regularity boost the performance of our proposed framework? In the following, we will first introduce the experimental setups including datasets, baselines and implementation details, and then present results as well as discussions.

\subsection{Datasets and Data Preparation.}

We use three benchmark datasets in our experiments: (i) \textbf{MovieLens-1M}: This dataset contains users' ratings (1-5) on movies. We treat samples with ratings greater than 3 as positive samples and samples with ratings below 3 as negative samples. Neutral samples with rating equal to 3 are dropped; (ii) \textbf{Criteo}: This dataset has 45 million users’ clicking records on displayed ads. It contains 26 categorical feature fields and 13 numerical feature fields; (iii) \textbf{Avazu}: This dataset has 40 million clicking records on displayed mobile ads. It has 22 feature fields spanning from user/device features to ad attributes. We follow the common approach to remove the infrequent features and discretize numerical values. First, we remove the infrequent feature values  and treat them as a single value ``\textit{unknown}", where threshold is set to $\{10, 4\}$ for Criteo, Avazu respectively. Second, we convert all numerical values in into categorical values by transforming a value $z$ to $\log^2(\text{int}(z))$ if $\text{int}(z)>2$ and to $\text{int}(z)-2$ otherwise, which is proposed by the winner of Criteo
Competition \footnote{\url{https://www.csie.ntu.edu.tw/~r01922136/kaggle-2014-criteo.pdf}}. Third, we randomly split all training samples into 80\% for training, 10\% for validation, and 10\% for testing. 

\subsection{Baselines}

We compare our proposed pruning method \texttt{HAM} with several baseline methods below.

\begin{itemize}[leftmargin=1em]
    \item \textbf{Uniform}: The method assigns an uniform embedding size for all feature fields;
    \item \textbf{Auxiliary Mask Pruning}: We compare our proposed method with other common approaches applying auxiliary mask to the embeddings. To be specific, we select three representative types of the auxiliary mask listed in Table \ref{tab:mask} and apply the same one-shot NAS framework as described in Algorithm \ref{alg:framework}:
    \begin{itemize}[leftmargin=1.5em]
    \item[(i)] \texttt{SAM}: the embeddings are masked directly by auxiliary weights $0\leq\alpha\leq1$, which is equivalent to \texttt{DNIS} \citep{cheng2020differentiable}; 
    \item[(ii)] \texttt{SAM-GS}: the auxiliary mask is obtained by the Gumbel-sigmoid function of auxiliary parameters $\alpha\in (0,1)$: 
    $$\quad \text{Sigmoid}[(\log(\frac{\alpha}{1-\alpha})+ \log(\frac{u}{1-u}))/\lambda],~ u\sim \text{Uniform(0,1)}.$$
    \item[(iii)] \texttt{HAM-p}: the mask is generated by Bernoulli random variables with parameters $p$'s when forwarding the network. The gradient with respect to $p$'s is calculated by STE
    \end{itemize}
    In the retraining stage, we keep the embedding components with top-$s$ auxiliary weights for fair comparisons.
    \item \textbf{AutoDim} \citep{zhao2020memory}: the state-of-the-art NAS-based (\textit{non-masking}) method in the literature, which outperforms a list of search methods \citep{chen2020differentiable, ginart2021mixed, zhao2020autoemb, joglekar2020neural}; see Section 3.4 in \citep{zhao2020memory}.
\end{itemize}

\subsection{Implementation Details}

 \textbf{Base Model architecture.} We adopt four representative CTR-prediction models in the literature: FM \citep{rendle2010factorization}, DeepFM \citep{guo2017deepfm}, AutoInt \citep{song2019autoint}, DCN-V2 \citep{wang2021dcn}, as our base models to compare their performance. For FM, DeepFM, AutoInt, we add an extra embedding vector layer between the feature interaction layer and the embedding layer as proposed in \citep{ginart2021mixed}. Linear transformations (without bias) are applied to mixed-dimension embeddings so that the transformed embedding vectors are of the same size, which is set as $16$ in our experiments. This is not required for DCN-V2 due to the bit-wise crossing mechanism. In the pretraining stage, the base dimension $d_j$ for each feature field $j$ is chosen as $\min(16, C_j)$ where $C_j$ is the field dimension. 

\textbf{Optimization.} We follow the common setup in the literature to employ Adam optimizer with the learning rate of $10^{-3}$ to optimize model parameters in all three stages, and use SGD optimizer to optimize auxiliary weights in the search stage. To fairly compare different \texttt{AMP} methods, the number of search epochs is fixed as 10, and the learning rates of auxiliary parameters are chosen from the search grid $\{1, 10^{-1}, 10^{-2}, 10^{-3}\}$, and the temperature of the sigmoid function is chosen from the search grid $\{1, 10^{-1}, 10^{-2}\}$. To obtain the best results of each method, we pick $10^{-2}$ as the learning rate of SGD for \texttt{SAM}, \texttt{SAM-p}, and \texttt{HAM-p}, $10^{-3}$ for our method \texttt{HAM}. In \texttt{HAM}, the initial value of auxiliary weights $\epsilon = 0.01$, and the hyperparameter $\mu=5\times 10^{-5}$. For the orthogonal regularity, we employ $\lambda \mathcal{R}(\sV)$ with $\lambda = 10^{-3}$ for training models on \texttt{MovieLens-1M}, and use the pair-wise cosine similarities $\lambda \mathcal{R}(\bar{\sV})$ with $\lambda = 10^{-6}$ on \texttt{Avazu}. We use PyTorch to implement our method and train it with mini-batch size 2048 on a single 16G-Memory Nvidia Tesla V100.

\subsection{Performance Comparison}

We next present detailed comparisons with other methods. We adopt AUC (Area Under the ROC Curve) on test datasets to measure the performance of models and the number of embedding parameters to measure the memory usage.

\begin{figure*}[h]
\centering
\includegraphics[width=0.9\linewidth]{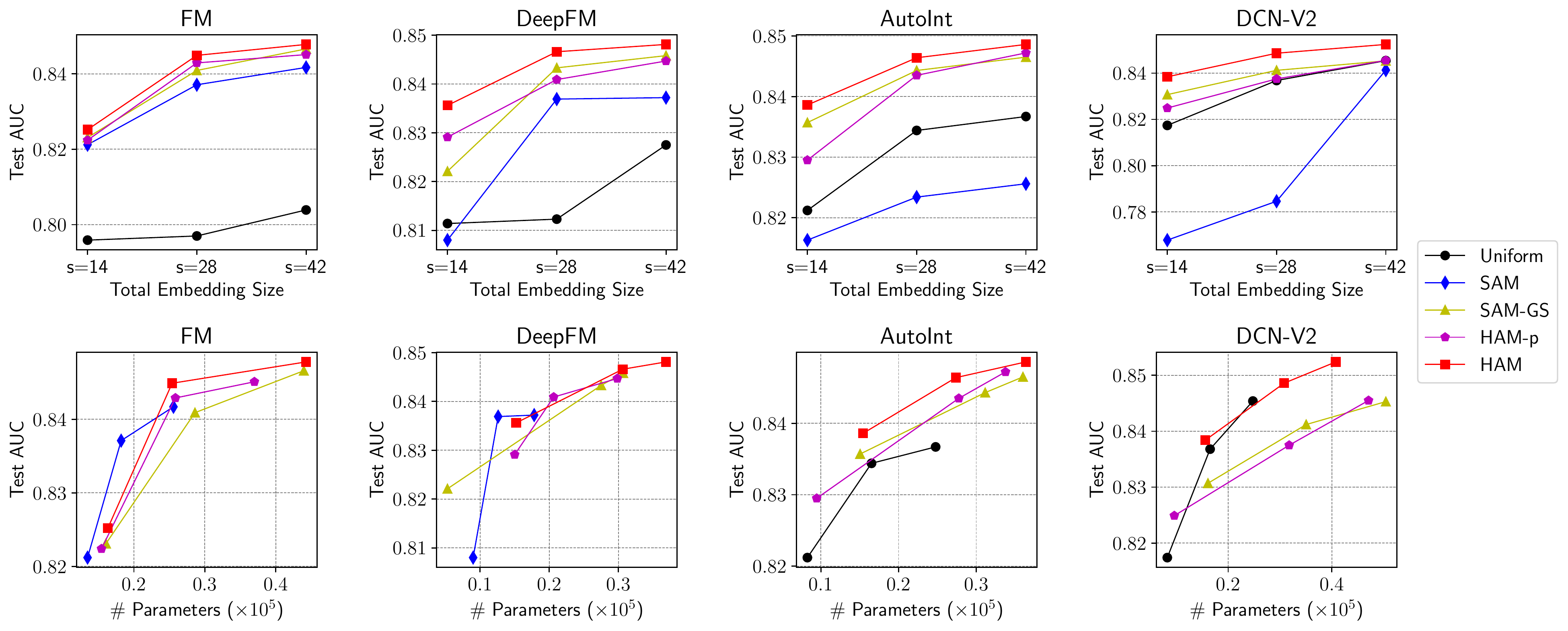}
\vskip -1em
\caption{ Performance comparison of different auxiliary mask pruning methods on MovieLens-1M. For each method, three measurements from left to right are reported with the target total embedding size $s = 14, 28, 42$ respectively.} 
\label{fig:ml-1m-amp}
\end{figure*}

\textbf{Comparison with other AMP Methods.} We first compare our proposed AMP methods to other common AMP methods listed in Table \ref{tab:mask} on \texttt{MovieLens-1M}. To fairly compare the performance, we not only consider comparing the test AUC under the same target total embedding size $s$ but also take the number of model parameters into account. In Figure \ref{fig:ml-1m-amp}, we report the average values of test AUC under the same total embedding size and the best results among 10 independent runs in terms of AUC and plot \textit{Test AUC - \# Parameter curve} on \texttt{MovieLens-1M}. For each method, three measurements are provided from left to right with the target total embedding size $s = 14, 28, 42$ respectively. We drop the \textit{Test AUC - \# Parameter curve} of the method with worst performance at the bottom. We observe that: \textbf{(a)} fixing the target total embedding size, we can tell that our method \texttt{HAM} outperforms all other methods on four base models with regard to the value of test AUC. As illustrated in \textit{Test AUC - \# Parameter curves}, with the same budget of total embedding size, \texttt{HAM} tends to assign larger sizes to informative feature fields with larger vocabulary sizes compared with other methods. This should be attributed to the gradient dynamics described in Section \ref{sec:gradient-dynamics} which helps to identify the important embedding components. As a result, \texttt{HAM} obtains models with higher AUC under the same total embedding size; \textbf{(b)} it is interesting to observe that \texttt{SAM} also assign larger embedding sizes to relatively un-informative features comparing to other methods with the same total embedding size, which verify the findings in \citep{wang2021rethinking} that the magnitudes of auxiliary weights may not indicate the importance; \textbf{(c)} \texttt{HAM} exhibits its superiority, not only on the vector-wise feature crossing models (FM, DeepFM, AutoInt), but also on the bit-wise feature crossing model (DCN-V2) where \texttt{Uniform} serves as a strong baseline. The performance of \texttt{HAM} should be credited to the hard selection caused by indicator functions. With the deterministic binary mask, the masked model evaluated in the search stage is equivalent to the selected model in the retraining stage, as illustrated in Table \ref{tab:mask}. As a result, we conclude that \texttt{HAM} exhibits stable performance on all four base models while \texttt{SAM}, \texttt{SAM-GS}, and \texttt{HAM-p} suffer from instability and suboptimality in some circumstances.

\begin{figure}[H]
\begin{center}
\includegraphics[width=0.9\linewidth]{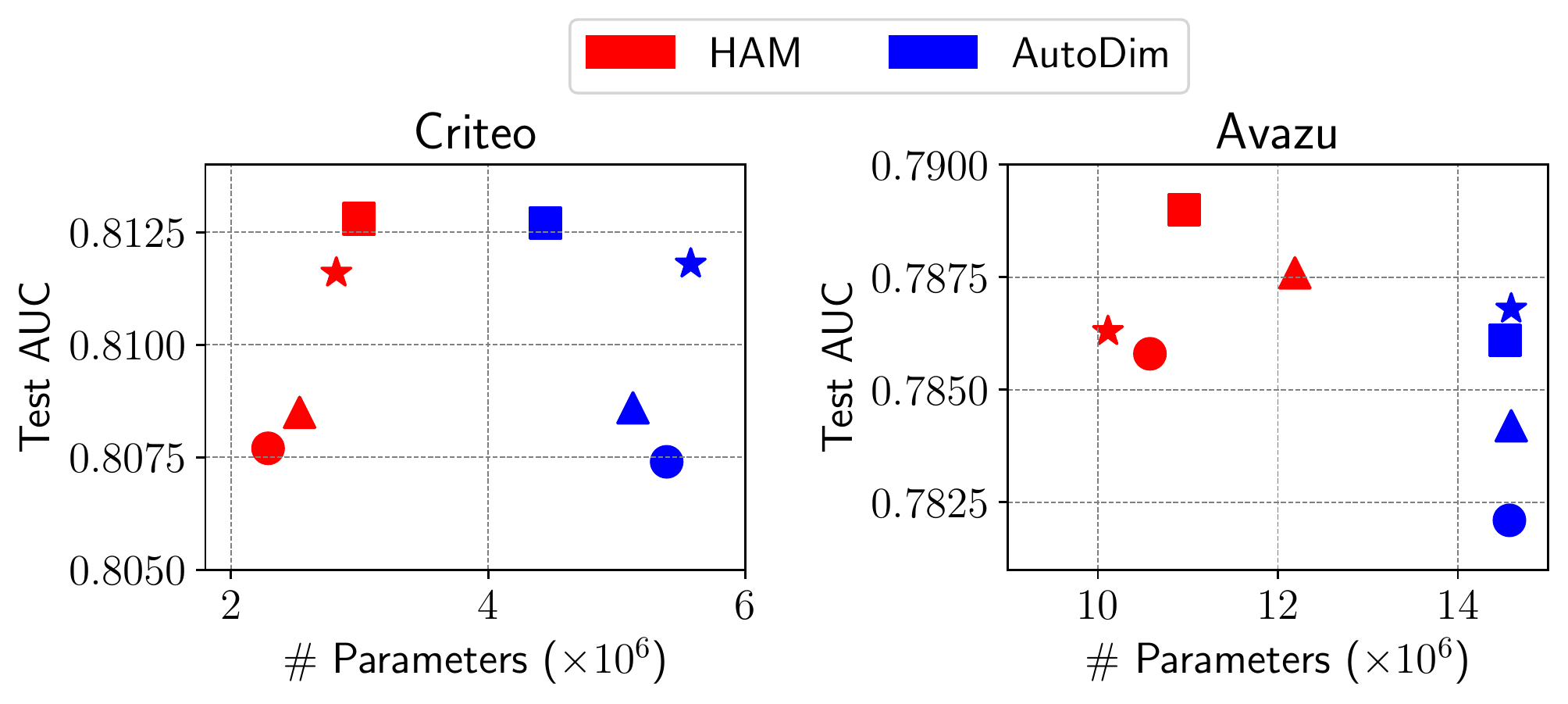}
\end{center}
\vskip -1em
\caption{Performance comparison between HAM and AutoDim. $\circ$ - FM; $\bigtriangleup$ - DeepFM; $\star$ - AutoInt; $\square$ - DCN-V2.}
\label{fig:autodim}
\end{figure}

\textbf{Comparison with AutoDim.} We also compare our method with state-of-art NAS-based method \texttt{AutoDim} with $\{2,4,8\}$ as the candidate embedding size  on \texttt{Criteo} and \texttt{Avazu} datasets in Figure \ref{fig:autodim}. In our method, the total embedding size $s$ is set to be $90, 50$ for Criteo and Avazu respectively. Our method outperforms \texttt{AutoDim} by finding models with comparably higher AUC and smaller size. In addition, from computational perspectives, our approach \texttt{HAM} has several advantages over \texttt{AutoDim}: \textbf{(a)} \texttt{HAM} has a larger search space due to the structural masking mechanism. For each feature, its candidate embedding size range from 0 to its base dimension $d_j$. On the contrary, \texttt{AutoDim} requires manually pre-specifying a set of candidate embedding sizes for each feature; \textbf{(b)} \texttt{HAM} is more computationally efficient. We emphasize that \texttt{AutoDim} introduce extra space and time complexity by the operations of lifting, batch normalization, and aggregation for each candidate size while \texttt{HAM} only requires extra element-wise product between the binary mask and embedding vectors. Moreover, \texttt{HAM} can output multiple models of different total embedding size given the same pre-trained model, whereas \texttt{AutoDim} requires pretraining the model once more if changing the candidate embedding size.

\begin{figure}[h]
\begin{center}
\includegraphics[width=0.8\linewidth]{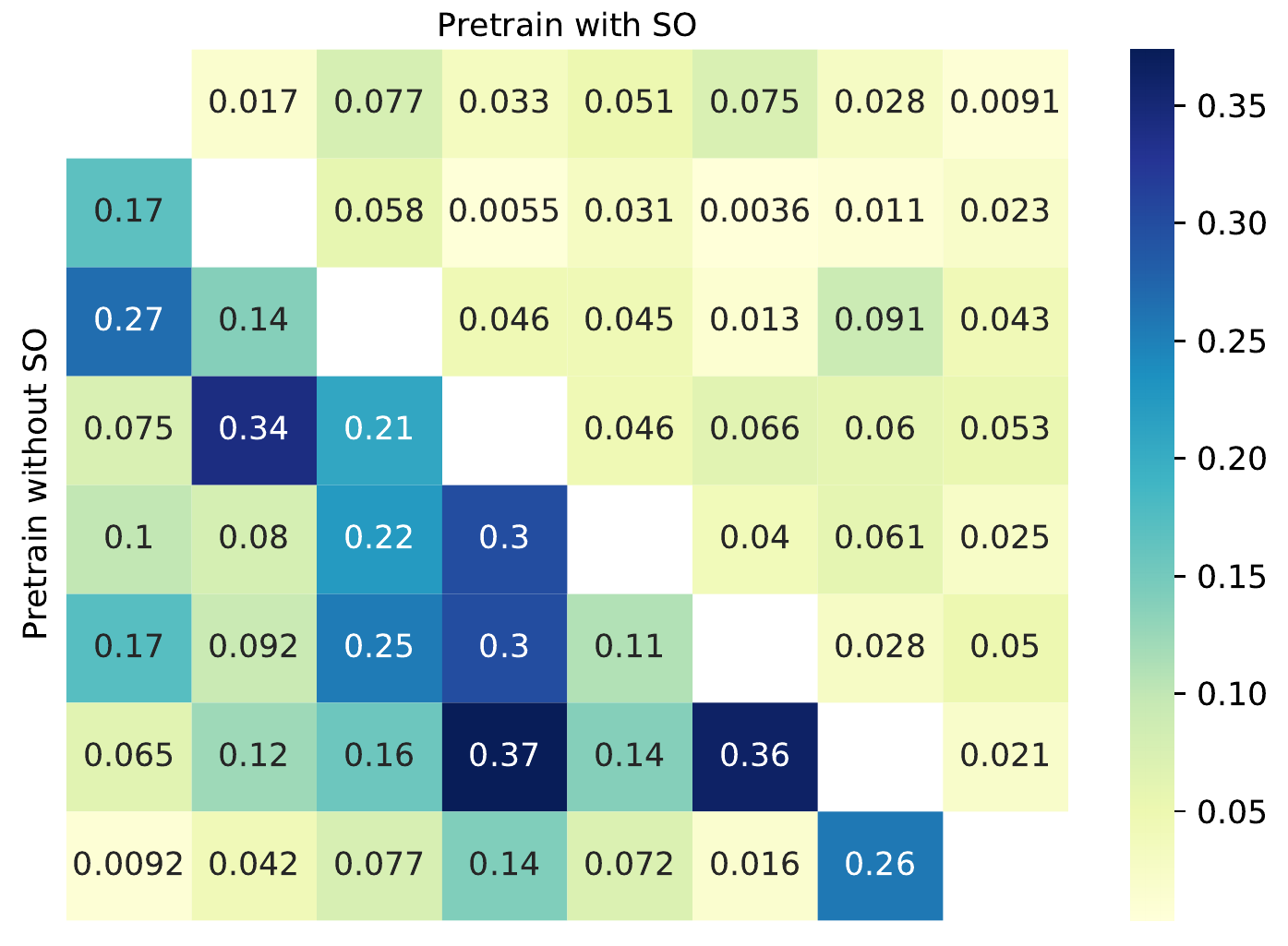}
\end{center}
\vskip -1em
\caption{Cosine similarities between column vectors from the embedding table of \textit{genre} when pretraining DCN-V2 on MovieLens-1M with and without SO.}
\label{fig:so}
\end{figure}

\begin{table}[h]
\small
\renewcommand\arraystretch{1}
\caption{The test AUC gain by Algorithm \ref{alg:framework} with SO under different target embedding sizes on MovieLens-1M and Avazu}
\vskip -0.5em
\label{tab:so-diff}
\begin{tabular}{>{\centering}p{0.12\linewidth}|>{\centering}p{0.08\linewidth}|>{\centering}p{0.14\linewidth}>{\centering}p{0.14\linewidth}
>{\centering}p{0.14\linewidth}
>{\centering\arraybackslash}p{0.14\linewidth}}
\toprule
 \multirow{2}{*}{\bf Dataset} &  \multirow{2}{*}{\bf Size $s$}  & \multicolumn{4}{c}{\bf Test AUC} \\ \cline{3-6}
 & & {\bf{FM}}  & {\bf{DeepFM}}  & {\bf{AutoInt}}  & {\bf{DCN-V2}}  \\
  \hline
 \multirow{4}{*}{\texttt{ML-1M}} & 14 &+.0006  &\textbf{+.0029}  &  +.0013 & +.0012 \\
   & 28  & \textbf{+.0018} & \textbf{+.0034}  & \textbf{+.0018} &   \textbf{+.0027} \\
   & 42 & \textbf{+.0036}  & \textbf{+.0033}  &  +.0009 &   \textbf{+.0037}  \\
  \hline
 \multirow{2}{*}{\texttt{Avazu}}   & 44 & \textbf{+.0034}  & \textbf{+.0039} &  \textbf{+.0026}  &  \textbf{+.0031}  \\
   & 66  & \textbf{+.0039} & \textbf{+.0044}  &  \textbf{+.0020} &   \textbf{+.0033} \\
\bottomrule
\end{tabular}
\vskip 0.1em
\textit{The results are obtained by averaging 10 runs and the bold font indicates statistically significant under two-sided t-test ($p < 0.05$)}.
\end{table}

\textbf{On the Orthogonal Regularity.} We further analyze the effect of the orthogonal regularity proposed in our framework. In Figure \ref{fig:so}, we visualize the cosine similarities between column vectors from the embedding of \textit{genre} when pretraining DCN-V2 on \texttt{MovieLens-1M} with and without \texttt{SO} where the embedding size is set as $8$. It is clear to see that \texttt{SO} helps to reduce the correlations within those embedding column vectors. Moreover, we compare the performance of models selected by our proposed Algorithm \ref{alg:framework} with the one without \texttt{SO} in Table \ref{tab:so-diff}. Statistical significant gains in test AUC can be observed for \texttt{MovieLens-1M} and \texttt{Avazu} on all base models while the training time per epoch with \texttt{SO} only increases by about 1.6 times compared to the time without \texttt{SO}.

\subsection{Further Discussions and Future Work}

\textbf{Multi-stage Search.} \label{sec:initial-size} As the initial base dimension for each feature fields it tunable, we observe a gain, $\sim$0.3\%, in test AUC from preliminary experiments on \texttt{MovieLens-1M} by searching via \texttt{HAM} with a smaller initial size $=8$. This observation confirms the claim made in recent works \citep{ci2021evolving} that enlarging search space is unbeneficial or even detrimental to existing NAS methods. To address this issue, our method can also be adapted to design a multi-stage search strategy with sufficiently larger initial size and stage-wisely prune embeddings with decreasing target embedding size $s$.

\textbf{Comparison with PEP.} To justify the advantages of our proposed framework, we also  about the most recent Plug-in Embedding Pruning (PEP) \citep{liu2021learnable} method. As introduced before, PEP is an unstructured pruning method which retains a large number of zero parameters and requires the technique of sparse matrix storage. On the contrary, our method can not only prune the embedding columns explicitly but also reduce the parameters in the dense layers while PEP cannot do. For example, \texttt{HAM} ($s=28$) can obtain a lighter DCN-V2 model on \texttt{MovienLens-1M} with only $\sim$8\% number of parameters (3,281/40,241) in the dense layers comparing to the model with a uniform embedding size 16.

\textbf{Comparison with SSEDS.} During the revisions of this paper, we noticed that a concurrent work \cite{qu2022single} also proposes to prune embeddings column-wisely as well as row-wisely using indictor functions. Our work is different from theirs in two aspects: (i) we focus on field-wise search by pruning nearly orthogonal embedding column vectors; (ii) we use a gradient-descent based method in the search stage to solve the bilevel problem on the training and validation set; the gradient dynamics enable us to re-activate (un-mask) masked embeddings while SSEDS directly mask all components with small gradient magnitudes. Comparing to SSEDS and incorporating our method into SSEDS is our future work.


\bibliographystyle{ACM-Reference-Format}
\bibliography{acmart}


\end{document}